\pdfoutput=1

\documentclass{article}

\usepackage{microtype}
\usepackage{graphicx}
\usepackage{subcaption}
\usepackage{booktabs}

\usepackage{hyperref}

\usepackage[accepted]{icml2026}

\usepackage{amsmath}
\usepackage{amssymb}
\usepackage{mathtools}
\usepackage{amsthm}

\usepackage[capitalize,noabbrev]{cleveref}

\usepackage{array}
\usepackage{tabularx}
\usepackage{ragged2e}
\newcolumntype{Y}{>{\RaggedRight\arraybackslash}X}

\icmltitlerunning{Scientific Knowledge Consolidation for AI-Driven Computational Physics}

\makeatletter
\renewcommand{\Notice@String}{Accepted by the ICML 2026 Workshop on AI for Physics (AI4Physics@ICML 2026).}
\makeatother

\begin{document}

\twocolumn[
  \icmltitle{From Experiments to Expertise: Scientific Knowledge\\Consolidation for AI-Driven Computational Physics}

  \begin{icmlauthorlist}
    \icmlauthor{Haonan Huang}{pton}
  \end{icmlauthorlist}

  \icmlaffiliation{pton}{Department of Physics, Princeton University, Princeton, NJ 08540, USA}

  \icmlcorrespondingauthor{Haonan Huang}{hnhuang@princeton.edu}

  \icmlkeywords{AI for physics, scientific reasoning, tool-augmented agents, structured memory, density functional theory, condensed matter, anomalous Hall conductivity}

  \vskip 0.3in
]

\printAffiliationsAndNotice{}

\begin{abstract}
While large language models (LLMs) have transformed AI agents into proficient executors of computational materials science, performing a hundred simulations does not make a researcher. What distinguishes research from routine execution is the progressive accumulation of knowledge --- learning which approaches fail, recognizing patterns across systems, and applying understanding to new problems. However, the prevailing paradigm in AI-driven computational science treats each execution in isolation, largely discarding hard-won insights between runs. Here we present QMatSuite, an open-source platform closing this gap. Agents record findings with full provenance, retrieve knowledge before new calculations, and in dedicated reflection sessions correct erroneous findings and synthesize observations into cross-compound patterns. In benchmarks on a six-step quantum-mechanical simulation workflow, accumulated knowledge reduces reasoning overhead by 67\% and improves accuracy from 47\% to 3\% deviation from literature --- and when transferred to an unfamiliar material, achieves 1\% deviation with zero pipeline failures.
\end{abstract}

\section{Introduction}

Large language model agents can now autonomously plan, execute, and interpret computational physics simulations, in many cases matching the proficiency of trained researchers on simulation tasks of various complexity~\cite{boiko2023coscientist,bran2024chemcrow,ghafarollahi2025atomagents,zou2025elagente,gustin2025agente,perez2026agente,kumar2026agente,bai2026agente,pham2026chemgraph,wang2025dreams,liu2025vaspilot,liu2025masgent}. The execution problem --- can AI run quantum-mechanical simulations correctly? --- is largely solved. Meanwhile, the broader AI community has seen landmark systems like the `AI Scientist'~\cite{natureaiscientist} successfully automate entire research lifecycles within machine learning environments.

Yet realizing this vision in the physical sciences exposes a fundamental barrier: physical execution is not a frictionless, rapid-feedback loop. A researcher who performs a hundred calculations over six months does not simply accumulate a hundred results. They internalize technical nuances, distinguish physical reality from artifacts, and distill cumulative observations into general principles. This progressive transformation of experience into expertise is what current AI systems lack. Existing agents start each session without access to findings from previous sessions: cross-session knowledge is either architecturally absent, limited to session-scoped scratchpads that reset between runs, or consists of static rules curated by human experts~\cite{zou2025elagente,pham2026chemgraph,wang2025dreams}. RAG over calculation logs provides a partial remedy but lacks mechanisms for quality validation, knowledge abstraction, and traceable provenance~\cite{lewis2020rag}. Recent surveys have identified experiential accumulation and consolidation as a critical frontier for AI agents broadly~\cite{hu2025agentmemory}; in computational physics specifically, structured memory for multi-step derivation and verification remains an open problem.

This gap will not close by building smarter models. The issue is structural: an agent session lasts minutes to hours, a physics research program unfolds over months to years. Bridging this requires infrastructure --- a platform where findings from one session are available to the next, where those findings can be reviewed and corrected against physical reasoning, and where accumulated observations can be synthesized into higher-order understanding.

We present QMatSuite, an open-source computational physics platform built to provide this infrastructure. We validate it across 135 autonomous solid-state calculations spanning six material categories and 98 molecular geometry optimizations --- using two different AI models and three simulation engines --- and through a controlled learning-curve experiment on the anomalous Hall conductivity (AHC) of bcc iron, a transport property that requires a six-step pipeline coupling two simulation codes and physical judgment at multiple steps. Accumulated knowledge transforms both the efficiency and the character of the agent's work, and transfers cleanly to an unfamiliar ferromagnet. We further show that knowledge consolidation --- synthesis of individual findings into cross-compound patterns --- requires dedicated reflection sessions separate from task execution, paralleling the cognitive rhythm of human physics research. The platform supports 15 simulation engines and connects to any AI model via the Model Context Protocol~\cite{anthropic2024mcp}, decoupling accumulated physics knowledge from both the computational engine that produced it and the AI model that will use it.

\section{Platform design and scale validation}

\begin{figure*}[!tbp]
\centering
\includegraphics[width=\textwidth]{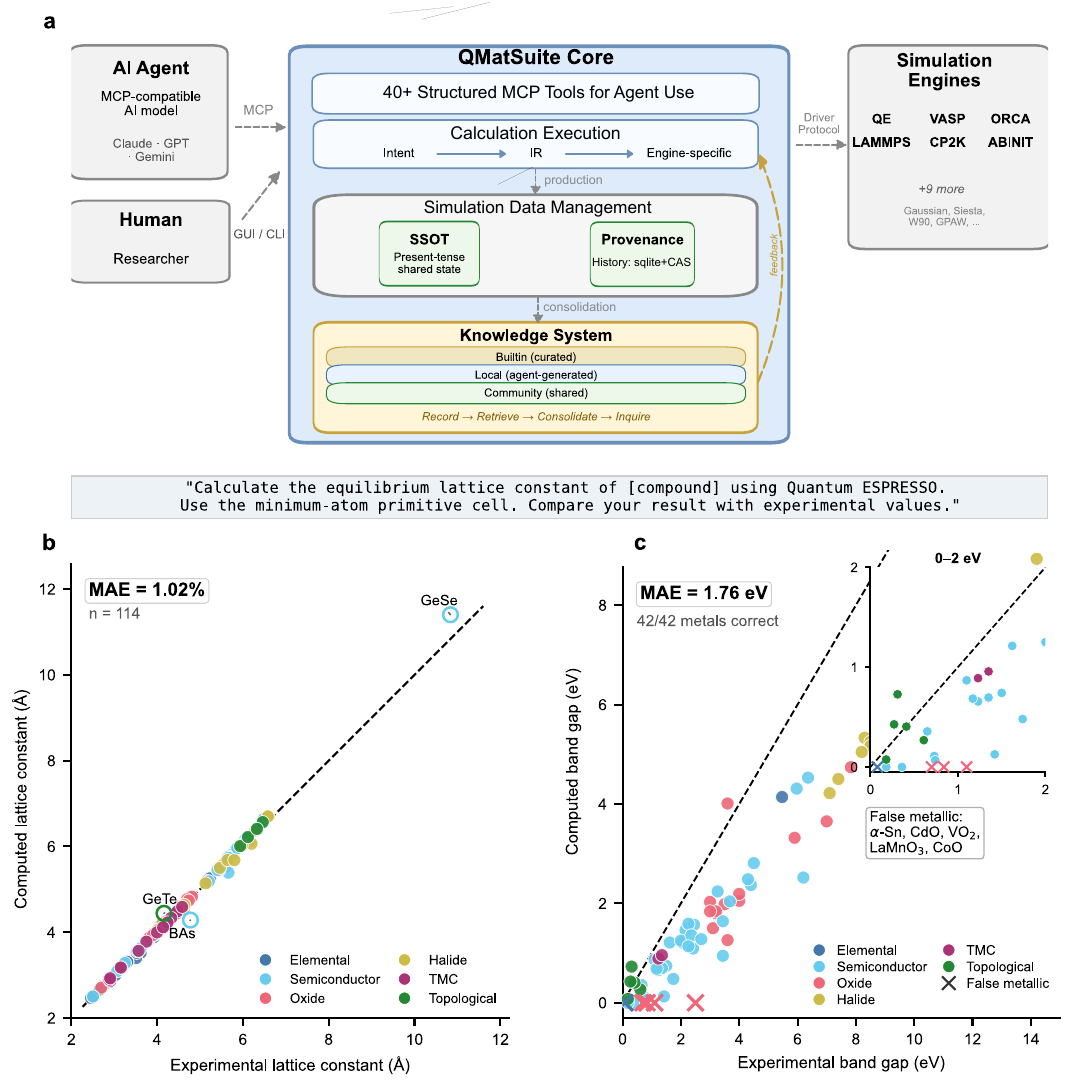}
\caption{\textbf{Platform architecture and scale validation.}
\textbf{a}, QMatSuite architecture showing symmetric access by AI agents (via MCP) and human researchers (via GUI) to the shared core, which contains the knowledge system, provenance tracking, present-tense shared state, and engine dispatch layer.
\textbf{b}, Computed versus experimental lattice constants for 114 materials (MAE = 1.02\%).
\textbf{c}, Computed versus experimental band gaps for 68 non-metallic compounds (MAE = 1.76~eV); 42/42 metals correctly identified; 5 correlated insulators predicted metallic (known PBE limitation, marked $\times$). Inset: 0--2~eV region.
}
\label{fig:platform}
\end{figure*}

QMatSuite is built on three core pillars. First, it abstracts simulation complexity through engine-agnostic structured tools (\texttt{set\_parameters}, \texttt{run\_calculation}, \texttt{get\_results\_summary}) that translate high-level agent calls into engine-specific inputs for 15 simulation codes (Appendix Table~\ref{edtab:competitors}); both AI agents (via MCP) and human researchers (via a desktop GUI) access the same core through a unified public API (Fig.~\ref{fig:platform}a). Second, end-to-end provenance tracks the entire history from raw inputs to final insights, creating an auditable chain in the spirit of reproducible workflow infrastructures~\cite{huber2020aiida,jain2015fireworks,wilkinson2016fair}. Third --- and central to this work --- the platform provides a persistent scientific memory system. Knowledge entries are organized in a graded hierarchy: \textit{findings} record observations from individual calculations (``PBE overestimates the GaAs lattice constant by 1.6\%''), \textit{patterns} synthesize regularities across multiple findings (``PBE overestimation scales with atomic mass across III-V compounds''), and \textit{principles} encode general rules. The knowledge base comprises curated best practices (read-only), agent-generated insights that accumulate across sessions (read-write), and community knowledge packs; this work focuses on the agent-generated pipeline.

Because calculation execution consumes the agent's full attention, the platform does not rely on the agent spontaneously pausing to manage knowledge. Instead, it embeds reminders at natural workflow junctures --- tool preambles prompt the agent to search prior knowledge before configuring a new calculation; post-execution returns prompt error-recovery recording; and results summaries prompt logging of numerical outcomes. These lightweight nudges make knowledge bookkeeping a natural byproduct of the calculation workflow rather than a separate task (Appendix~\ref{app:methods}); their efficiency is tested directly below. All experiments are driven by short natural-language prompts specifying only scientific intent (reproduced verbatim in each figure banner); agents receive no guidance on computational parameters, software syntax, or workflow structure.

\paragraph{Scale validation.}
To validate the platform at scale, we deployed an autonomous agent (Claude Sonnet 4.6) via QMatSuite's MCP interface to drive structural relaxations and band structure calculations in Quantum ESPRESSO (QE)~\cite{giannozzi2009qe,giannozzi2017qe}. Testing across 135 diverse materials spanning six categories (elemental metals and semiconductors, binary semiconductors, simple oxides, halides, layered transition metal compounds, and topological insulators) (Fig.~\ref{fig:platform}b,c) yielded an 85.2\% autonomous completion rate (115 materials). Success rates scaled with structural complexity, from 100\% in simple metals to 62--65\% in topological insulators and layered transition metal compounds. Notably, the 20 incomplete cases resulted exclusively from timeouts rather than crashes or erroneous outputs.

Lattice constants for 114 materials agree with experiment at 1.02\% mean absolute error (MAE), consistent with established PBE benchmarks~\cite{PhysRevB.79.085104}. Band gaps for 68 non-metallic compounds show 1.76~eV MAE, dominated by the expected PBE underestimation. All 42 metals were correctly identified; five strongly correlated insulators ($\alpha$-Sn, CdO, VO$_2$, LaMnO$_3$, CoO) were predicted metallic --- known semilocal-DFT failures, not agent errors~\cite{mori2008bandgap,cohen2008limitations}. The agent recorded a median of 3 insights per material, searching prior knowledge 4.2 times and recording 3.2 times on average --- confirming that the platform's nudge-based knowledge bookkeeping integrates naturally into the calculation workflow without explicit user instructions.

\paragraph{Cross-engine and cross-model validation.}
To verify engine and model agnosticism, we deployed GPT~5.4 with ORCA~6.1.1~\cite{neese2020orca} for 98 molecular geometry optimizations (Appendix Fig.~\ref{edfig:orca}), of which 91 completed successfully. Bond lengths for the reference-backed raw-ORCA-converged subset agree with spectroscopic references at MAE $=$ 0.0069~\AA{} (0.52\%), and bond angles at MAE $=$ 0.51$^\circ$, consistent with established hybrid-DFT geometry benchmarks~\cite{buehl2006geometries,mardirossian2017thirty}. The seven final failures comprise two agent-declared failures despite converged raw outputs, three missing or non-terminated last-output cases, and two explicit last-attempt error terminations.

\section{Knowledge transforms a complex workflow}

\subsection{The learning curve}

The 135-material validation demonstrates competence on routine computational tasks. The more demanding test is whether accumulated knowledge helps when task complexity exceeds what pre-training alone can handle --- when parameters must be coordinated across multiple codes, failure modes are undocumented, and physical judgment is required at multiple steps. We chose the anomalous Hall conductivity (AHC) of bcc iron as the testbed: a quantum-mechanical transport property~\cite{RevModPhys.82.1539} whose computation requires a six-step pipeline coupling two simulation codes (QE and Wannier90~\cite{pizzi2020wannier90}), with a large parameter space where many settings require non-trivial physical reasoning. Even experienced practitioners require significant trial and error; the literature value is approximately $-751$~S/cm~\cite{yao2004ahc}.

\begin{figure*}[!tbp]
\centering
\includegraphics[width=\textwidth]{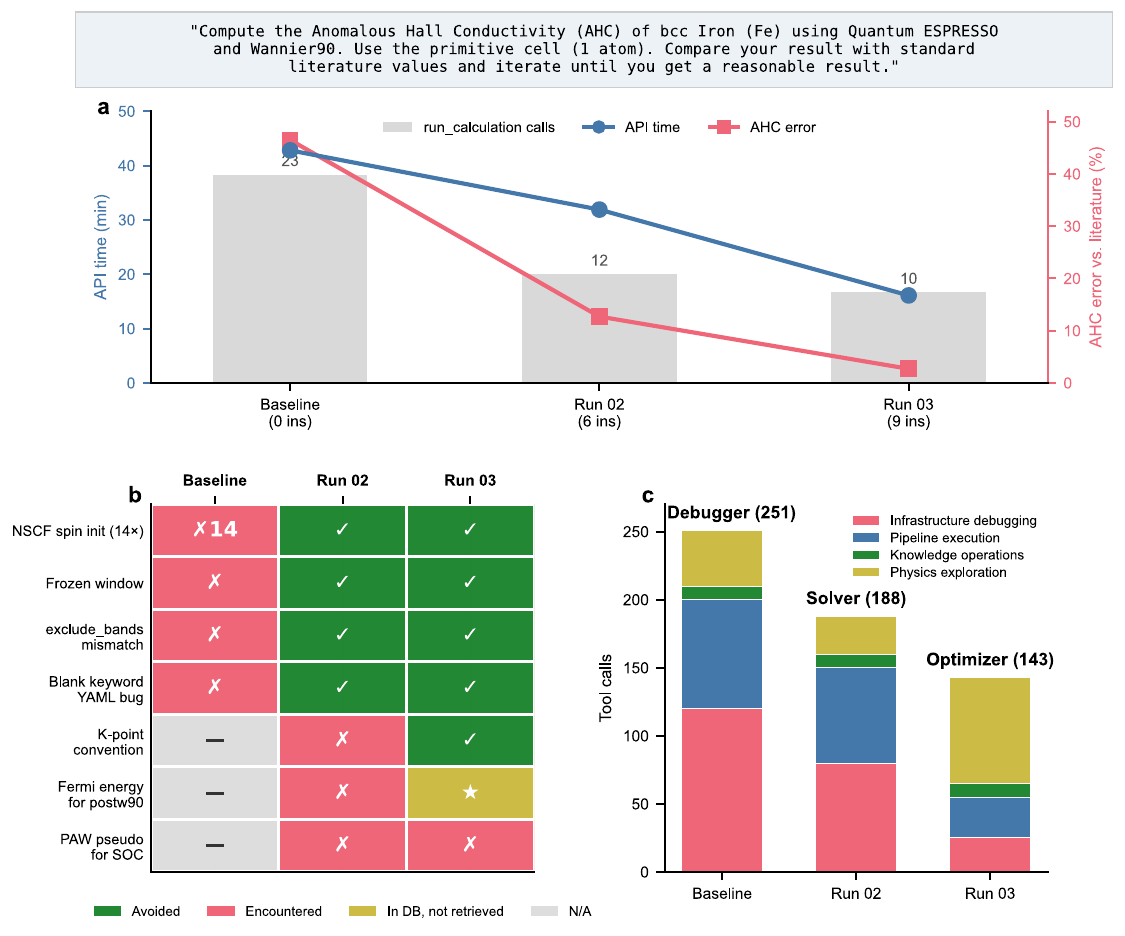}
\caption{\textbf{Knowledge transforms a complex workflow.}
\textbf{a}, Fe AHC learning curve across three runs with 0, 6, and 9 accumulated insights. Gray bars: \texttt{run\_calculation} calls; blue line: API reasoning time; red line: AHC error versus literature.
\textbf{b}, Episode avoidance matrix showing which pitfalls each run encountered (red) versus avoided via knowledge (green). Yellow: insight in database but not retrieved; gray: not applicable.
\textbf{c}, Tool call composition shifting from infrastructure debugging (``Debugger'') to physics exploration (``Optimizer'').
}
\label{fig:hero}
\end{figure*}

We ran the identical task three times in sequence (Fig.~\ref{fig:hero}), each as a fresh session with a fresh project directory and the same model (Claude Opus 4.6). The sole variable carried between runs was the knowledge database: the baseline started empty, the second run inherited 6 insights from the first, and the third inherited 9.

Every metric improves monotonically (Fig.~\ref{fig:hero}a). API reasoning time --- which isolates cognitive effort from simulation compute --- dropped from 42.8 to 16.1 minutes (62\% reduction), total tool calls from 251 to 143, and pipeline execution attempts from 23 to 10. AHC accuracy improved from $-1100$~S/cm (46.5\% error) to $-846.2$~S/cm (12.7\%) to $-771.5$~S/cm (2.7\%).

The dominant time saving traces to a single insight. In the baseline, the agent spent around 3 hours and 14 pipeline attempts diagnosing why the computed AHC was zero --- ultimately discovering that QE's non-self-consistent (NSCF) step requires explicit non-zero \texttt{starting\_magnetization}; otherwise, it silently defaults to a non-magnetic state, producing zero Berry curvature and zero AHC. This behavior --- physically counterintuitive because NSCF reuses converged charge densities --- is neither clearly documented in standard manuals nor represented in training data. The finding was recorded as insight~\#2. In both subsequent runs, the agent retrieved this insight before its first calculation and applied the fix proactively, entirely eliminating the debugging episode (Fig.~\ref{fig:hero}b).

Three observations establish the knowledge database, rather than pre-training knowledge, as the causal variable (Appendix Table~\ref{edtab:retrieval}). First, the identical model without the database required 14 attempts and 3.5 hours to discover the spin initialization requirement. Second, the third run failed to retrieve an available insight and consequently re-encountered the corresponding error --- demonstrating that retrieval, not latent model knowledge, mediates transfer. Third, the monotonic improvement across three runs with identical model weights but increasing database content is consistent only with the database as the independent variable.

\subsection{Cognitive liberation}

Total wall time does not monotonically decrease --- and this tells the most interesting story. The baseline agent finished at 6.2 hours, most of it consumed by infrastructure debugging; it never had the opportunity to question whether its result was converged. The second run completed in under two hours but also did not attempt convergence optimization. The third run reached its first valid result ($-921$~S/cm, 23\% error) in 58 minutes --- then voluntarily spent an additional 19 hours on a systematic seven-iteration convergence study, varying disentanglement parameters and comparing uniform versus adaptive mesh refinement strategies. It discovered that adaptive refinement outperforms brute-force mesh densification at one-seventh the computational cost --- a genuine methodology insight that no previous run had the cognitive bandwidth to pursue.

The three runs thus exhibit a qualitative behavioral progression (Fig.~\ref{fig:hero}c and Appendix Fig.~\ref{edfig:timeline}): from debugger (the baseline, spending 70\% of tool calls on infrastructure), through solver (the second run, resolving known and novel errors to reach a valid result), to optimizer (the third run, freed from debugging to pursue systematic physical exploration). The accuracy improvement from 47\% to 3\% is a consequence of this liberation, not of the knowledge system directly optimizing parameters.

\subsection{Knowledge quality and self-correction}

Across three runs, the agent accumulated 15 insights (Appendix Table~\ref{edtab:insights}): 9 error-recovery findings (60\%), 3 result recordings (20\%), and 3 methodology insights (20\%) --- the last category appearing only in the third run, reflecting the behavioral shift toward physics exploration. Of the 15, 13 are fully correct, one is partially correct, and one contains correct data but an incorrect conclusion. This last case involves Wannier90's disentanglement parameter \texttt{dis\_froz\_max}, which sets the upper energy boundary of the ``frozen window'' --- the range within which Bloch states are included exactly in the Wannier representation~\cite{PhysRevB.65.035109}. For transport properties that depend on states near the Fermi level, this window should extend well above the Fermi energy to ensure accuracy of relevant bands. The agent's third run found that setting \texttt{dis\_froz\_max} to 17~eV (slightly below the Fermi energy of ${\sim}$17.5~eV) gave the best AHC value and recommended keeping it low --- prioritizing an accidental fit to literature over rigorous convergence validation (Fig.~\ref{fig:transfer}a).

In a dedicated review session, the same model operating in reflection mode identified this error (Fig.~\ref{fig:transfer}a). Applying convergence analysis to the agent's own parameter sweep data, it recognized that the recommended value was an unconverged outlier while higher values produced stable results. It further verified this against official tutorial input files retrieved from raw source repositories, which directly contradicted the calculation agent's recommendation. The insight was deprecated and replaced with a corrected version citing both the convergence data and the tutorial evidence. This self-correction required structured general guidance, but the agent finds answers itself (Appendix~\ref{app:review}).

\subsection{Cross-material transfer}

\begin{figure*}[!tbp]
\centering
\includegraphics[width=\textwidth]{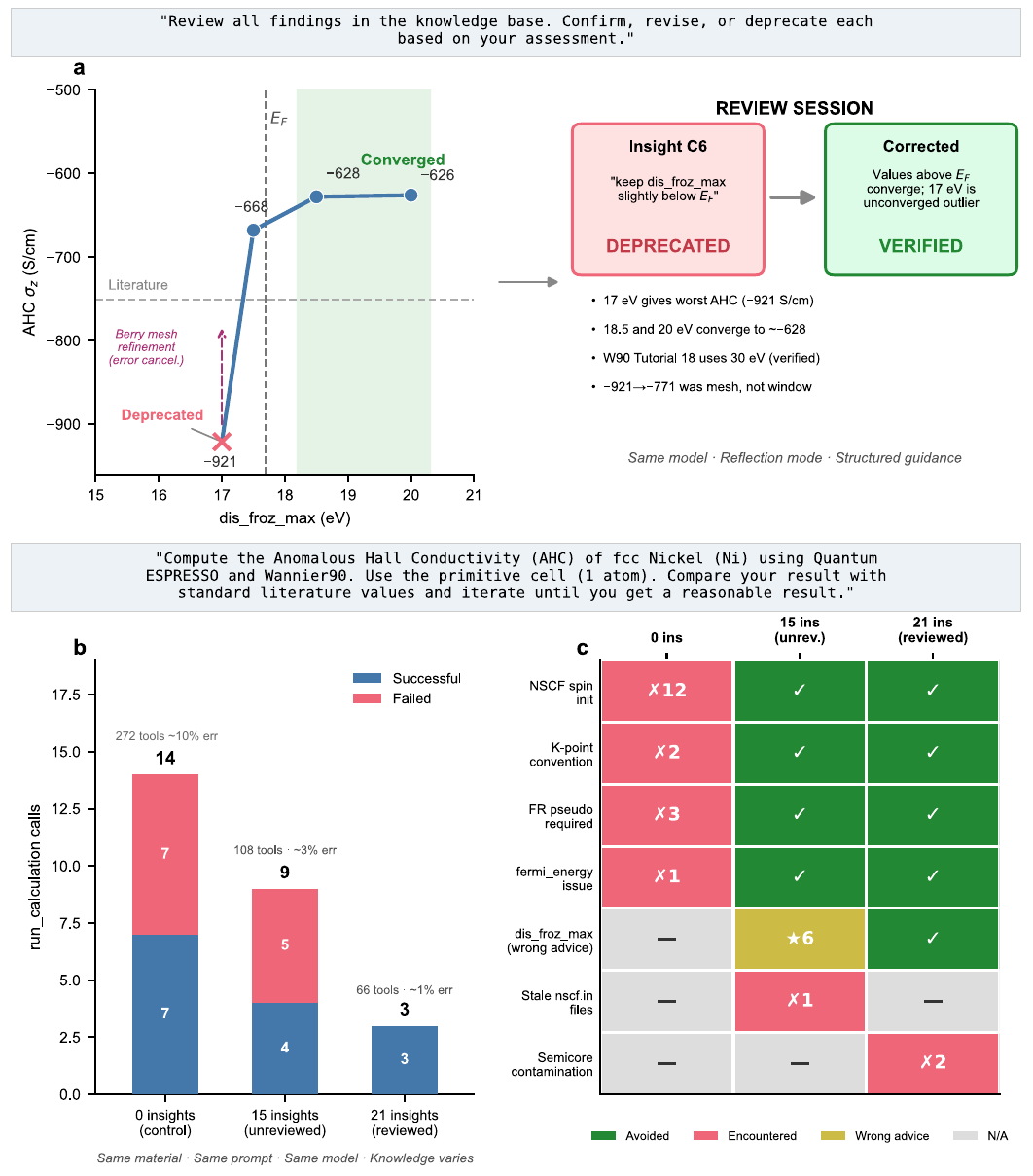}
\caption{\textbf{Knowledge self-corrects and transfers across materials.}
\textbf{a}, Left: convergence analysis showing the deprecated parameter recommendation (\texttt{dis\_froz\_max}~$=$~17~eV, red $\times$) as an unconverged outlier, with the upward arrow indicating that Berry mesh refinement coincidentally pulled the AHC toward the literature value (error cancellation). Right: the review session's deprecation reasoning and corrected replacement.
\textbf{b}, Ni AHC three-way comparison: 0 insights (14 calculations, 7 failures), 15 unreviewed insights (9 calculations, 5 failures), 21 reviewed insights (3 calculations, 0 failures).
\textbf{c}, Ni pitfall avoidance matrix showing progressive transfer; yellow star ($\bigstar$6) marks the \texttt{dis\_froz\_max} row where unreviewed knowledge provided incorrect advice requiring 3 extra iterations.
}
\label{fig:transfer}
\end{figure*}

To test whether knowledge generalizes beyond the material that produced it, we applied the reviewed iron knowledge database (21 insights) to a nickel AHC calculation --- a system with no prior entries in the database. The prompt was identical except for the element name (Fig.~\ref{fig:transfer}b,c).

The agent completed the Ni AHC in 66 tool calls with only 3 pipeline executions --- all successful --- achieving $|\sigma_z| = 2079$~S/cm (within 1.0\% deviation from the literature value of ${\sim}2073$~S/cm~\cite{wang2007fermi}). API reasoning time was 14.1 minutes. The agent proactively applied correct pseudopotentials, spin settings, and convergence strategies from iron knowledge.

A controlled three-way Ni comparison shed light on the effect of knowledge quality (Fig.~\ref{fig:transfer}b,c). The naive baseline agent (0 insights) required 14 pipeline executions --- 7 failures --- and 272 tool calls to reach ${\sim}10\%$ error. The agent with 15 unreviewed Fe insights reduced this to 9 executions but inherited an incorrect parameter recommendation, costing three additional iterations, but the agent can still recover from such knowledge poisoning. The reviewed 21-insight agent completed the task in 3 executions with zero failures.

The most revealing finding concerns what did \textit{not} transfer. Without a prior nickel calculation record, the 21-insight agent could not engage in recipe replication --- the behavior observed in all six same-material iron sessions (Appendix~\ref{app:review}), where a key parameter was copied verbatim from the highest-performing prior result despite being flagged as incorrect. For nickel, it instead reasoned from general principles, arriving at a physically correct parameter choice. Strikingly, the same agent achieved better accuracy on an unfamiliar material (1.0\%) than on the familiar one (16--23\% in same-material reruns). While the Ni result may partly reflect favorable error cancellation in the absence of a systematic convergence study, the contrast suggests that recipe availability can actively hinder principle-based reasoning.

\section{Knowledge consolidates into understanding}

\subsection{From findings to patterns}

The AHC experiments demonstrate that accumulated findings improve efficiency and accuracy on a single complex workflow. A natural next question is whether agents can move beyond individual findings to synthesize higher-order understanding --- recognizing regularities that span multiple calculations and materials. To test this, we ran a chain of 24 calculation sessions across 19 zinc-blende semiconductors, during which the agent accumulated 25 individual findings --- each a structured record of one calculation's outcome (e.g., ``PBE overestimates the GaAs lattice constant by 1.6\%''). No higher-order synthesis occurred during any task-execution session: across 398 execution sessions, 84.9\% received the platform's recording nudges, yet 95.3\% of all knowledge recordings occurred in end-of-session batches and zero pattern-grade entries --- cross-compound generalizations that synthesize multiple findings --- were produced (Appendix Fig.~\ref{fig:consolidation}b,c). Trace analysis reveals that agents informally recognized cross-compound trends during execution, yet never deposited these as higher-grade knowledge entries, because attention was fully consumed by the current calculation.

A dedicated reflection session with a 12-word prompt (``Review accumulated findings and summarize any patterns you see across compounds'') produced three quantitative patterns from 25 individual findings in under three minutes (Appendix Fig.~\ref{fig:consolidation}a). Pattern 1 identified that PBE lattice constant overestimation scales systematically with atomic mass across all three compound families (Group IV, III-V, II-VI), with quantitative error ranges tabulated per family --- a synthesis spanning 19 compounds that no single session could have produced. Pattern 2 distinguished quantitative band gap underestimation from qualitative topological failures in near-zero-gap systems. Pattern 3 established that variable-cell relaxation with semicore pseudopotentials requires kinetic energy cutoffs above 80~Ry, supported by 13 successful calculations with Pulay stress identified as the causal mechanism~\cite{francis1990finite}. Each pattern references its supporting findings by unique identifier, maintaining full provenance to source calculations. These are not trivial restatements --- they contain structured quantitative data, causal reasoning, and explicit scope boundaries.

An independent session using a less capable model (Claude Sonnet 4.6 instead of Opus 4.6) produced three patterns with identical themes from the same 25 findings, confirming that pattern structure is latent in the data rather than an artifact of a specific model.

This parallels a well-established observation in human cognition: synthesis does not occur during task execution but requires dedicated reflective activity~\cite{sio2009incubation,baird2012inspired}. Independently, recent work on experiential reinforcement learning has identified an analogous reflection-consolidation loop as necessary for durable improvement in AI agents~\cite{shi2026erl}. Our findings provide empirical evidence for this principle in a scientific computing context (Appendix Fig.~\ref{fig:consolidation}).

\subsection{From understanding to new inquiry}

As a proof of principle, we prompted the agent to test one of its patterns. It selected InP, predicted a PBE band gap of ${\sim}0.7$~eV by interpolating Pattern 2's underestimation trend, and computed 0.696~eV --- confirming the prediction. While this amounts to interpolation along a monotonic curve, it demonstrates that the full knowledge cycle --- from observation through consolidation to hypothesis-driven inquiry --- can close autonomously. Whether agents can design more physically creative experiments from richer knowledge bases remains an intriguing open question for future work.

\section{Discussion}

The experiments presented here demonstrate that persistent scientific memory transforms AI agent behavior beyond simple acceleration. The learning curve shows the most visible effect --- a 67\% reduction in reasoning overhead from baseline to cross-material transfer --- but the deeper transformation is qualitative. The runs exhibit a behavioral progression from debugger to solver to optimizer to transferable expert: the baseline agent spent its entire session on infrastructure; the knowledge-equipped agent on an unfamiliar material completed the task with zero pipeline failures.

Agent-generated knowledge is predominantly correct, demonstrating that the recording system produces reliable scientific documentation. Like any research output, it occasionally captures incorrect conclusions: the \texttt{dis\_froz\_max} recommendation was based on correct numerical data but drew the wrong inference from under-converged calculations. Critically, unreviewed incorrect knowledge has measurable cost: the 15-insight Ni agent required 3 extra iterations to override inherited bad advice, while the reviewed 21-insight agent encountered zero failures. This establishes knowledge review not as an academic exercise but as a practical necessity that measurably reduces wasted computation.

The knowledge system directly addresses the RAG limitations~\cite{lewis2020rag,liu2024lost,packer2023memgpt} identified in the Introduction, with our experiments providing direct evidence for each. Regarding correctness, unreviewed knowledge containing flawed recommendations results in performance penalty; the reviewed database eliminated this cost entirely. For compression, 25 individual findings were consolidated into 3 patterns in under three minutes; such abstraction renders retrieval and community sharing trivial due to the reduced data footprint. Finally, concerning provenance, when the review session deprecated the \texttt{dis\_froz\_max} recommendation, the system preserved the complete lineage from the erroneous conclusion back to the source data, ensuring the correction remained fully auditable and verifiable.

Cross-session memory for LLM agents is by now an established line of work: reflective and episodic memory~\cite{shinn2023reflexion,park2023generativeagents}, memory-augmented language models~\cite{packer2023memgpt}, skill and workflow memory~\cite{wang2023voyager,wang2024awm}, and a recent wave of dedicated agent-memory systems~\cite{xu2025amem,chhikara2025mem0,rasmussen2025zep,fang2025memp}. What none of these combine, and QMatSuite does, is three things at once: a graded, \textit{reviewable}, provenance-linked findings$\to$patterns$\to$principles hierarchy with explicit deprecation, in which unreviewed knowledge carries a \textit{measured} cost rather than an assumed one; a \textit{quantified} separation of execution from reflection, with consolidation empirically confined to dedicated review sessions; and a domain --- computational physics --- in which both the process and the outcome of a calculation have \textit{independent} ground truth. That independent ground truth is exactly what makes the error-cancellation and recipe-replication phenomena reported here observable at all, and it is out of reach for memory studied only on chat or question-answering benchmarks. The typed knowledge graph proposed in review --- with deprecation status as a first-class node property --- is a natural next step that could make recipe replication detectable structurally rather than behaviorally, and we see it as promising future work.

Agents exhibit distinct cognitive modes that have direct implications for system design. In execution mode, agents are goal-directed and resistant to mid-task reflection --- only 4.0\% of sessions show any mid-session knowledge recording despite 84.9\% receiving explicit nudges, and zero pattern-grade entries were produced across 398 sessions. In reflection mode, the same model performs convergence analysis, tutorial verification, and self-correction~\cite{park2023generativeagents,shinn2023reflexion}. Effective AI research programs should alternate between execution and reflection sessions, with tool design adapted to each mode.

The cross-material experiment reveals the essential role of knowledge quality, and that recipe replication --- copying parameters from the highest-performing prior result rather than reasoning about their applicability --- is an anti-pattern triggered by recipe availability. In all six same-material iron sessions, the agent selected the deprecated \texttt{dis\_froz\_max} value by extracting it from the best-performing result record, ignoring deprecation warnings. On unfamiliar nickel, the absence of a material-specific recipe forced the agent to reason from principles --- and it achieved comparable or better accuracy. Knowledge systems should encourage principle-level entries over numerical recipes.

QMatSuite demonstrates robust generality across engines (Quantum ESPRESSO, Wannier90, ORCA) and models (Claude and GPT). These results are directly relevant to physics-centric reasoning with LLM agents: structured, provenance-linked, reviewable memory is what converts a tool-augmented agent --- which can already run quantum-mechanical simulations correctly --- into one that performs multi-step derivation, convergence verification, and physical-principle reasoning across sessions.

Sufficiently capable models will not obviate this infrastructure: the bottleneck is session isolation, not reasoning capability; smarter models will produce higher-quality insights and sharper reviews, making the infrastructure more valuable, not less. The platform supports community-contributed, domain-specific knowledge packs~\cite{wilkinson2016fair}; persistent scientific memory, demonstrated here for condensed-matter calculations, represents a general requirement for AI-driven research across computational physics.

The reflection loop is not infallible. A review session could in principle consolidate a false pattern, promoting a spurious regularity to higher-grade knowledge. Because review verdicts are themselves recorded as reviewable entries and every entry retains its provenance chain back to the source calculations, such errors remain auditable and reversible, and that same provenance supports human-in-the-loop pruning of poisoned entries. How robustly the loop behaves when a weaker or open-weights model occupies the reflection role --- rather than the frontier model used here --- is untested, and we leave it to future work.

Several limitations bound these claims. The learning-curve evidence rests on a single sequential three-run chain, without repeated trials or variance estimates; the Fe$\to$Ni transfer stays within the 3$d$-ferromagnet family, a relatively mild out-of-distribution test; and we do not benchmark against long-context replay or raw-log RAG baselines, which we leave as direct comparisons for future work. The efficiency result is likewise best read as a gain in cognitive reallocation --- from debugging toward physical exploration --- rather than as reduced wall time.

\section*{Code and data availability}

QMatSuite is open-source and available at \url{https://github.com/QMatSuite/QMatSuite}. Complete experimental artifacts --- full prompts, agent traces, knowledge-database snapshots, and reviewed-insight tables --- are archived on Zenodo at \url{https://doi.org/10.5281/zenodo.20588420} under CC-BY 4.0.

\section*{Impact Statement}

This paper presents work whose goal is to advance AI-driven computational physics. Persistent scientific memory systems may accelerate physics research, with corresponding risks for the propagation of incorrect inherited knowledge when review practices are weak --- risks we discuss through the knowledge-quality and recipe-replication analyses above. The manuscript itself was written by human authors describing an AI-agent platform; we do not advocate for autonomous publication of physics results without human review.

\bibliographystyle{icml2026}
\bibliography{references}

\newpage
\appendix
\onecolumn

\section{Methods}
\label{app:methods}

\paragraph{Platform architecture.}
QMatSuite provides 40$+$ structured tools via the Model Context Protocol (MCP), organized into five categories: Project (initialization, structure management), Config (parameter setting, presets), Execute (calculation launch, monitoring), Analysis (results extraction, structure promotion), and Knowledge (search, record, review). Engine drivers translate structured tool calls into engine-specific input files. The platform uses a three-layer execution model: user/agent intent (through structured operations such as \texttt{set\_parameters}, \texttt{apply\_preset}, \texttt{run\_calculation}), an engine-agnostic intermediate representation, and engine-specific materialization into QE namelists, VASP INCAR tags, ORCA keyword blocks, or other engine-native inputs. Fifteen simulation engines are supported at varying integration depth: Quantum ESPRESSO~\cite{giannozzi2009qe,giannozzi2017qe}, VASP, ORCA, LAMMPS, CP2K, ABINIT, Gaussian, Siesta, Wannier90, GPAW, Psi4, PySCF, xTB, QMCPACK, and Yambo. All experiments in this work used Quantum ESPRESSO 7.5~\cite{giannozzi2009qe,giannozzi2017qe} with SSSP precision pseudopotentials~\cite{prandini2018sssp} for solid-state calculations and ORCA 6.1.1~\cite{neese2020orca} for molecular calculations.

\paragraph{Persistent scientific memory.}
The knowledge system comprises three tiers stored in SQLite databases with FTS5 full-text search indexing. The builtin tier contains curated best practices (read-only). The local tier stores agent-generated insights (read-write) in three grades: findings (individual observations), patterns (cross-system regularities), and principles (general rules). Each entry includes scope tags, source calculation references, and free-text reasoning. The community tier supports sharable domain-specific insight packs. All tiers are searched via a unified \texttt{search\_knowledge} tool with relevance ranking. Recording uses \texttt{record\_insight} with mandatory grade, scope, and reasoning fields. Review uses \texttt{review\_insight} with verdicts (confirmed, deprecated, promoted) and provenance links. Insights may reference other insights by ULID, enabling hierarchical knowledge structures.

\paragraph{Prompt design and nudge mechanism.}
All experiments used short natural-language task prompts specifying scientific intent only (e.g., ``Calculate the equilibrium lattice constant of [material]''); they contained no computational parameters, software syntax, or workflow structure. Experiment-level system preambles imposed workflow discipline (engine execution through structured tool calls, one active run at a time, explicit error-recovery recording) without prescribing chemistry-specific parameter choices. The nudge mechanism embedded reminders in tool response text at natural workflow junctures: a front-loaded reminder in \texttt{create\_calculation} to search prior knowledge before configuration; a post-recovery reminder in \texttt{run\_calculation} to record the error and its fix; a results-linked reminder in \texttt{get\_results\_summary} to record numerical findings; and a synthesis reminder in \texttt{list\_insights} (e.g., ``$N$ findings pending synthesis. Consider reviewing with \texttt{list\_insights(grade='finding')} when your current task is complete''). These nudges increased opportunity for knowledge capture but did not guarantee compliance.

\paragraph{AHC learning curve experiment.}
The anomalous Hall conductivity of bcc iron was computed using the Quantum ESPRESSO~$+$ Wannier90 six-step pipeline: SCF~$\to$~NSCF~$\to$~wannierprep~$\to$~pw2wannier90~$\to$~wannier90~$\to$~postw90. The agent was Claude Opus 4.6 with no timeout. The prompt was identical across all three runs. Each run used a fresh Claude session, a fresh project directory, and the same model. The sole variable was the knowledge database: Condition A started empty (0 insights), Run 02 inherited 6 insights, and Run 03 inherited 9. The builtin knowledge database was disabled (\texttt{QMS\_KNOWLEDGE\_BUILTIN=0}) to isolate local knowledge effects. To ensure the knowledge database was the sole information channel between sessions, web search tools were blocked, engine installation tutorial and example files were removed from disk, and session persistence was disabled. All agent traces were audited post-hoc to verify that no external information sources --- web, filesystem tutorials, or cached content from prior runs --- were accessed in any experimental session. The model's pre-training knowledge constitutes an uncontrolled stochastic baseline; the learning curve design isolates the database contribution from this baseline.

\paragraph{Cross-material transfer experiment (Ni).}
Three Ni AHC runs used the same prompt (substituting ``fcc Nickel (Ni)'' for the Fe element) and the same model (Claude Opus 4.6, no timeout). The baseline (0 insights) used an empty knowledge database with all leakage channels blocked; the trace was audited post-hoc to confirm no external information access. Run 26 (15 insights) used the unreviewed Fe knowledge database containing two incorrect parameter recommendations. Run 24 (21 insights) used the reviewed database with deprecated entries corrected. The literature reference for fcc Ni AHC is approximately 2000--2400~S/cm~\cite{yao2004ahc,wang2007fermi,fuh2011intrinsic,vzelezny2023high}.

\paragraph{Meta session and knowledge consolidation experiment.}
19 zinc-blende semiconductors underwent sequential relaxation and band structure calculations (24 sessions, Claude Opus 4.6) accumulating 25 findings in the local knowledge database. Two independent reflection sessions --- one using Claude Sonnet 4.6, one using Claude Opus 4.6 --- were then run with an identical 12-word prompt. Both models produced 3 patterns with identical themes, confirming that pattern structure is latent in the data rather than model-specific. A subsequent design experiment session (Claude Opus 4.6) used a 29-word prompt with no preamble, attempt limits, or tool instructions.

\paragraph{Nudge compliance analysis.}
Nudge compliance was measured across 398 execution sessions (270 from the 135-material benchmark, 128 from AHC and reflection experiments) by parsing all \texttt{trace.jsonl} files programmatically. Recording nudge delivery was identified by matching nudge phrases in tool response text. Each \texttt{record\_insight} call was located by event index within its session and classified by position (early $<$30\%, mid 30--70\%, late 70--90\%, end-of-session $>$90\%). Of 398 sessions, 338 (84.9\%) received at least one recording nudge; 16 (4.0\%) contained any mid-session \texttt{record\_insight} call; 95.3\% of all recordings occurred in the final 10\% of session events; and zero sessions produced pattern-grade entries. All 41 patterns in the knowledge base were produced exclusively in dedicated review sessions.

\paragraph{Cross-engine molecular validation.}
98 molecules spanning diatomics, triatomics, small polyatomics, organics, radicals, and transition metal complexes were computed using ORCA 6.1.1 with agent-selected methods (predominantly B3LYP/def2-TZVP-level hybrid DFT~\cite{weigend2005balanced,mardirossian2017thirty}). The agent was GPT 5.4. Experimental reference bond lengths and angles were compiled from NIST CCCBDB, the NIST Chemistry WebBook, Huber \& Herzberg spectroscopic constants, and primary spectroscopic literature.

\section{Platform comparison and retrieval evidence}

\begin{table}[h]
\centering
\small
\caption{Comparison of AI agent systems for computational science. Eight systems evaluated across capability dimensions relevant to knowledge-driven research. ``Intent-level tools'' means the agent expresses scientific purpose (e.g.\ \texttt{run\_calculation}) rather than writing engine-specific input files. Knowledge persistence refers to whether agent-generated findings survive across independent sessions.}
\label{edtab:competitors}
\renewcommand{\arraystretch}{1.25}
\resizebox{\textwidth}{!}{%
\begin{tabular}{@{}lllllllll@{}}
\toprule
\textbf{System} & \textbf{Venue} & \textbf{Input gen.} & \textbf{Param select.} & \textbf{Error recovery} & \textbf{Cross-session knowledge} & \textbf{Knowledge review} & \textbf{Engines} & \textbf{Open src} \\
\midrule
El Agente Q & \textit{Matter} 2025 & ORCA files & \checkmark & \checkmark & Static rules$^*$ & --- & 1 & No \\
ChemGraph & \textit{Commun. Chem.} 2026 & ASE API & Partial & Limited & None & --- & 6 & Yes \\
DREAMS & arXiv 2025 & QE via ASE & \checkmark & \checkmark & Session-scoped & --- & 1 & Yes \\
AtomAgents & \textit{PNAS} 2025 & Pre-built fn & Partial & Limited & None & --- & 1 & Yes \\
VASPilot & \textit{Chin.\ Phys.\ B} 2025 & VASP files & Hybrid & \checkmark & RAG (wiki) & --- & 1 & Yes \\
Masgent & arXiv 2025 & VASP files & Defaults & Preventive & None & --- & 1 & Yes \\
Aitomia & arXiv 2025 & \checkmark & \checkmark & \checkmark & None & --- & ML & Partial \\
\textbf{QMatSuite} & \textbf{This work} & \textbf{Intent tools} & \textbf{\checkmark} & \textbf{\checkmark} & \textbf{Persistent, graded} & \textbf{\checkmark} & \textbf{15} & \textbf{Yes} \\
\bottomrule
\end{tabular}%
}
\vspace{4pt}
{\footnotesize $^*$El Agente Q's ``hierarchical memory'' consists of static rules manually curated by human experts; episodic memory is explicitly disabled in the published implementation.}
\end{table}

\begin{table}[h]
\centering
\small
\caption{Insight retrieval and application across Fe AHC runs. Matrix showing which insights from the knowledge database were retrieved and applied in each run. $\bigstar$~marks an insight present in the database but not retrieved, causing the agent to re-encounter the corresponding error. This case (insight B2 in Run 03) provides causal evidence that retrieval --- not latent model knowledge --- mediates transfer: the same model with the same pre-training failed to avoid an error whose solution was available but not found by BM25 search.}
\label{edtab:retrieval}
\renewcommand{\arraystretch}{1.2}
\begin{tabular}{@{}clcccc@{}}
\toprule
& & \multicolumn{2}{c}{\textbf{Run 02}} & \multicolumn{2}{c}{\textbf{Run 03}} \\
\cmidrule(lr){3-4} \cmidrule(lr){5-6}
\textbf{ID} & \textbf{Content (brief)} & \textbf{Retrieved} & \textbf{Applied} & \textbf{Retrieved} & \textbf{Applied} \\
\midrule
A1 & Frozen window fix & \checkmark & \checkmark & --- & --- \\
A2 & \texttt{starting\_mag} (CRITICAL) & \checkmark & \checkmark & \checkmark & \checkmark \\
A3 & Semicore exclusion & --- & Set indep. & \checkmark & \checkmark \\
A4 & \texttt{exclude\_bands} trap & \checkmark & Avoided & --- & --- \\
A5 & AHC $-1100$ (prior result) & \checkmark & Planning & \checkmark & Planning \\
A6 & Blank keyword bug & \checkmark & Avoided & --- & --- \\
\midrule
B1 & K-point convention & \multicolumn{2}{c}{\textit{(Recorded this run)}} & \checkmark & \checkmark \\
B2 & \texttt{fermi\_energy} required & \multicolumn{2}{c}{\textit{(Recorded this run)}} & $\bigstar$ \textbf{Not retr.} & Hit error \\
B3 & AHC $-846$ (improved) & \multicolumn{2}{c}{\textit{(Recorded this run)}} & \checkmark & Planning \\
\bottomrule
\end{tabular}
\vspace{4pt}
{\footnotesize $\bigstar$ Insight B2 was in the database but none of Run 03's four search queries matched it. The agent independently resolved the error and recorded a redundant insight (C2). This demonstrates that BM25 retrieval is imperfect --- having knowledge in the database does not guarantee it will be found.}
\end{table}

\section{Knowledge inventory for the Fe AHC learning curve}
\label{app:insights}

Across three runs, the agent accumulated 15 insights. In summary: 13 are fully correct, 1 is partially correct (A5: correct calculation, incomplete convergence attribution), and 1 contains correct data but an incorrect conclusion (C6: \texttt{dis\_froz\_max} recommendation based on unconverged outlier). The load-bearing deprecated entry is reproduced in full below as the worked example for the self-correction claim; the complete insight table is available in the archived artifacts (\url{https://doi.org/10.5281/zenodo.20588420}).

\begin{table}[h]
\centering
\small
\caption{Deprecated insight C6 (full content), the load-bearing example for the knowledge self-correction claim. Entry contains correct sensitivity data but draws an incorrect conclusion: \texttt{dis\_froz\_max~$=$~17} produced the best AHC ($-921$~S/cm) through error cancellation (Berry mesh refinement compensated for unconverged Wannier functions); higher values (18.5, 20) converge to $\sim\!-628$~S/cm. Deprecated and replaced in a dedicated review session via convergence analysis and tutorial verification.}
\label{edtab:insights}
\begin{tabularx}{\textwidth}{@{}lY@{}}
\toprule
\textbf{Field} & \textbf{Value} \\
\midrule
ID & C6 \\
Source run & Run 03 (iron AHC, Claude Opus 4.6) \\
Grade & Finding (method) \\
Scope tags & wannier90, AHC, disentanglement \\
Original content & ``\texttt{dis\_froz\_max}$=$17~eV (slightly below $E_F=17.5$~eV) gave the best AHC result ($-921$~S/cm baseline, $-771.5$~S/cm with adaptive mesh); keep this value low.'' \\
Supporting data & Sweep: \texttt{dis\_froz\_max}$=$17$\to$$-921$; 17.5$\to$$-668$; 18.5$\to$$-628$; 20$\to$$-626$~S/cm \\
Correctness & Wrong conclusion (data correct) \\
Review verdict & Deprecated \\
Replacement & ``\texttt{dis\_froz\_max} must be well above $E_F$ to ensure convergence of states near the Fermi level; the best literature agreement at 17~eV arose from error cancellation with Berry-mesh refinement. Recommended: $\geq$20~eV, verified against Wannier90 Tutorial~18 (\texttt{dis\_froz\_max$=$30}).'' \\
\bottomrule
\end{tabularx}
\end{table}

\textbf{Counts by run.} Baseline (Run A): 6 entries (A1 frozen-window fix, A2 NSCF \texttt{starting\_magnetization} critical requirement, A3 semicore exclusion, A4 \texttt{exclude\_bands} trap, A5 $-1100$~S/cm result, A6 blank-keyword crash). Run 02: 3 entries (B1 k-point convention, B2 \texttt{fermi\_energy} required, B3 $-846$~S/cm result). Run 03: 6 entries (C1 projections string form, C2 redundant \texttt{fermi\_energy}, C3 FR pseudopotential required for SOC, C4 $-772$~S/cm adaptive-mesh result, C5 adaptive-refinement vs brute-force cost comparison, C6 deprecated method entry above). Three entries are methodology insights --- all appearing in Run 03, reflecting the behavioral shift toward physics exploration.

\section{AHC trace analyses}
\label{app:ahc}

\paragraph{Cross-run summary for the Fe learning curve.}
\begin{center}
\small
\begin{tabular}{@{}lccc@{}}
\toprule
Metric & Cond. A (0) & Run 02 (6) & Run 03 (9) \\
\midrule
Final AHC (S/cm) & $-1100$ & $-846.2$ & $-771.5$ \\
Error vs.\ literature & 46.5\% & 12.7\% & 2.7\% \\
API time (min) & 42.8 & 31.9 & 16.1 \\
Wall time (min) & 372.4 & 113.3 & 1211.3 \\
Total tool calls & 251 & 188 & 143 \\
\texttt{run\_calculation} calls & 23 & 12 & 10 \\
\texttt{search\_knowledge} calls & 4 & 3 & 4 \\
\texttt{record\_insight} calls & 6 & 3 & 6 \\
\bottomrule
\end{tabular}
\end{center}

Every cognitive-effort metric improved monotonically as the database grew. Wall time did not, because Run 03 spent the majority of its time on a deliberate convergence study rather than infrastructure debugging.

\paragraph{Condition A: discovery from a blank database.}
Condition A lasted 21{,}761~s and terminated by timeout after 251 tool calls. The session began with empty \texttt{search\_knowledge} and \texttt{list\_insights} results. The first major challenge was pseudopotential selection: the auto-resolution routine returned a PAW file inappropriate for spin--orbit coupling, forcing the agent to locate and stage a fully relativistic pseudopotential manually. The next obstacle arose at Wannierization: with \texttt{num\_wann~$=$~18}, a frozen window placed too high included too many states; the agent diagnosed this from the eigenvalue spectrum and lowered the frozen-window boundary. The decisive diagnostic came from comparing SCF and NSCF outputs: the SCF calculation was magnetic, but the NSCF output reported a non-magnetic spin--orbit calculation. The agent eventually inferred that \texttt{starting\_magnetization(1)} must also be present in the NSCF input. This single discovery consumed roughly 3.5~h and 14 pipeline attempts in the baseline, and it became the most important transferable error-recovery finding in the learning curve. The final AHC value was $-1100$~S/cm, overestimated by 46.5\%.

\paragraph{Run 02: inherited knowledge removes the zero-AHC trap.}
Run 02 began with six findings from Condition A in the local database. Before the first calculation was launched, the agent retrieved and explicitly planned around four of them. This inheritance removed the entire zero-AHC episode from the session. The dominant bottleneck instead became a new interoperability bug: a k-point convention mismatch between Quantum ESPRESSO and Wannier90. Roughly 40\% of the trace was spent exploring explicit crystal-format k-point lists, stale input-file collisions, and shifted mesh conventions. The eventual fix was to supply explicit crystal k-points folded into the $[-0.5, 0.5)$ convention for the NSCF step. After a further correction to supply the Fermi energy to the transport post-processing stage, the pipeline completed cleanly at $-846.2$~S/cm. Run 02 thus provides the first direct causal demonstration of persistent knowledge: a previously discovered failure mode disappeared entirely, while a new and previously unseen failure mode became the dominant source of effort.

\paragraph{Run 03: knowledge liberates the agent for convergence analysis.}
Run 03 inherited nine findings and retrieved five unique insights before the first calculation. Those retrieved findings directly shaped the initial setup: explicit NSCF k-points, explicit NSCF magnetization, a lower frozen-window floor, a denser SCF mesh. The run still encountered three problems (FR pseudopotential bug in the automatic resolver; incorrect serialization of the Wannier90 projection field; omission of the Fermi energy in the final transport stage), all corrected and recorded. Once the first successful pipeline returned $-921$~S/cm, the character of the session changed --- the agent was no longer infrastructure-bound and used the rest of the session for a seven-iteration convergence study:

\begin{center}
\small
\begin{tabularx}{\textwidth}{@{}cYcY@{}}
\toprule
Iter. & Key change & $\sigma_z$ (S/cm) & Interpretation \\
\midrule
1 & Baseline, uniform Berry mesh $100^{3}$ & $-921$ & First end-to-end converged pipeline, still overestimated \\
2 & \texttt{dis\_froz\_max~$=$~20}~eV & $-626$ & Upper frozen window moved into the converged regime \\
3 & \texttt{dis\_froz\_max~$=$~18.5}~eV & $-628$ & Same converged branch as 20~eV \\
4 & \texttt{dis\_froz\_max~$=$~17.5}~eV & $-668$ & Intermediate regime \\
5 & No frozen window & $-3.4$ & Berry curvature collapses \\
6 & Uniform Berry mesh $200^{3}$ & $-864$ & Brute-force refinement helps, at very high cost \\
7 & Adaptive refinement + \texttt{use\_ws\_distance} & $-771.5$ & Best result, 2.7\% error \\
\bottomrule
\end{tabularx}
\end{center}

The database did not directly encode the final adaptive-refinement recipe. Rather, it removed enough repeated debugging that the agent could spend the remainder of the session on systematic physical exploration --- the mechanistic basis for the ``cognitive liberation'' claim.

\paragraph{Cross-material transfer to Ni.}
The three-way Ni comparison isolates the effects of knowledge quantity and quality:

\begin{center}
\small
\begin{tabularx}{\textwidth}{@{}Ycccccc@{}}
\toprule
Run & Start insights & Final AHC (S/cm) & Error & \texttt{run\_calc} calls & Failed runs & Tool calls \\
\midrule
0 insights & 0 & $+$2289 & $\sim$10\% & 14 & 7 & 272 \\
15 insights (unreviewed Fe) & 15 & $+$2040 & $\sim$3\% & 9 & 5 & 108 \\
21 insights (reviewed Fe) & 21 & $+$2079 & 1.0\% & 3 & 0 & 66 \\
\bottomrule
\end{tabularx}
\end{center}

The reviewed 21-insight agent reached $|\sigma_z|~=~2078.6$~S/cm in three successful launches and 14.1~min of API time. Importantly, it did not copy the Fe-specific \texttt{dis\_froz\_max~$=$~17}~eV recipe. Because the database contained Fe-specific findings but no Ni-specific ``best-result'' record, the agent instead reasoned from general principles: use a fully relativistic pseudopotential, keep NSCF explicitly magnetic, use explicit crystal k-points, and inspect the eigenvalue spectrum to choose the frozen-window placement. It therefore began at \texttt{dis\_froz\_max~$=$~20}~eV, diagnosed residual semicore contamination after the second run, widened the outer window, and converged in the third iteration.

\begin{figure}[h]
\centering
\includegraphics[width=\textwidth]{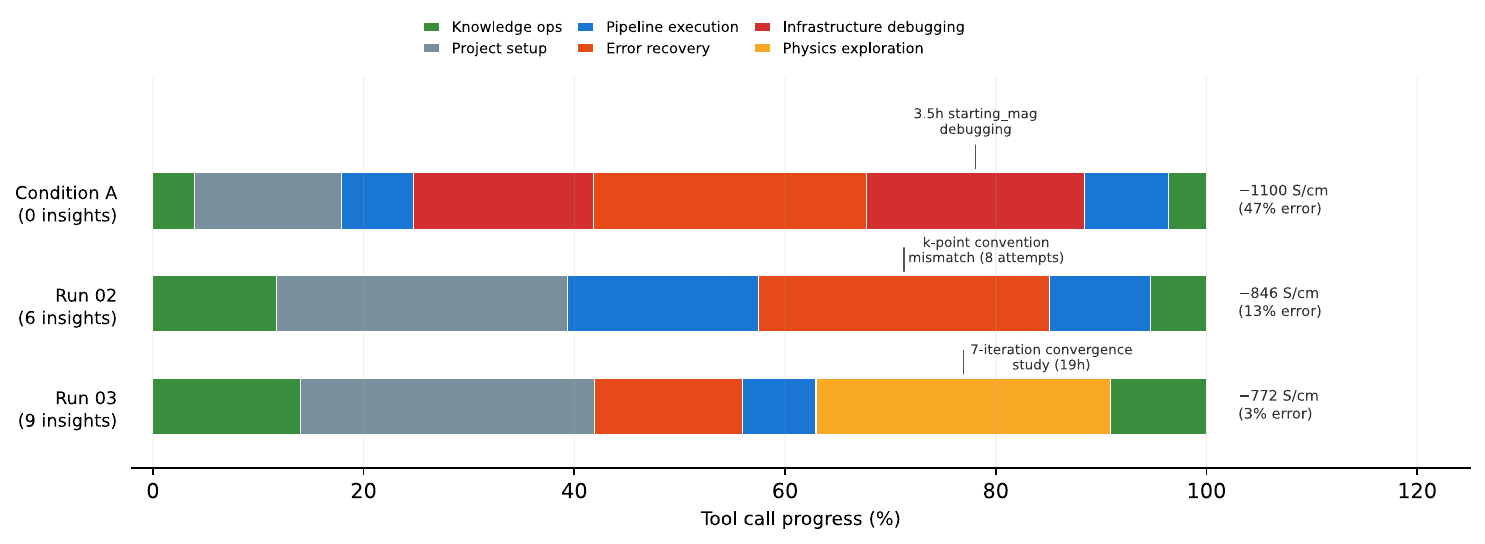}
\caption{\textbf{Temporal structure of agent activity across three Fe AHC runs.}
Gantt-style visualization of tool call composition, showing the shift from infrastructure debugging to physics exploration as knowledge accumulates. \textbf{Condition A} (0 insights): dominated by debugging and error recovery, with 3.5~h spent diagnosing why the computed AHC was zero (the \texttt{starting\_magnetization} issue). \textbf{Run 02} (6 insights): proactively avoided the \texttt{starting\_mag} pitfall but encountered a novel k-point convention mismatch requiring 8 attempts. \textbf{Run 03} (9 insights): resolved all setup issues within the first 80 tool calls, then devoted the remainder to a voluntary 7-iteration convergence study --- a qualitative shift from ``getting it to work'' to ``understanding the physics.''}
\label{edfig:timeline}
\end{figure}

\section{Review-session journey and the knowledge-delivery gap}
\label{app:review}

\paragraph{The target finding.}
The review experiment was designed around a subtle but scientifically important failure mode: a finding containing correct numerical data but an incorrect conclusion. Insight C6 correctly recorded the Fe AHC sensitivity sweep but concluded that \texttt{dis\_froz\_max} should be kept slightly below the Fermi level because the final adaptive-refinement run at 17~eV happened to give $-771.5$~S/cm, close to the literature value. In reality, that agreement arose from error cancellation between an unconverged frozen-window choice and later adaptive Berry-mesh refinement.

\paragraph{Layered defenses.}
The review experiment did not converge on a single perfect prompt. Different preamble layers resolved different cognitive failure modes, and the resulting progression required a layered defense stack. The scientific lesson is not prompt sensitivity in the abstract but that claim-testing review required composing multiple distinct mechanisms:

\begin{center}
\small
\begin{tabularx}{\textwidth}{@{}p{3.6cm}Y Y@{}}
\toprule
Preamble layer & Failure mode addressed & Mechanism \\
\midrule
Citation URL $+$ verbatim excerpt & Hallucinated verification & Forces genuine external retrieval rather than memory-only verification \\
``Separate data from conclusions'' & Correct data, wrong inference & Makes the model test the conclusion rather than only the data record \\
``Search recommendations directly'' & Topic-level instead of claim-level retrieval & Encourages dedicated search for the recommendation itself \\
Step~0 PREPARATION (fetch raw input files) & Evidence-discovery gap & Moves the search target from prose pages to raw calculation inputs \\
``No middle ground'' & Confirmed-with-caveat rationalization & Removes hedged verdicts for partially wrong findings \\
\bottomrule
\end{tabularx}
\end{center}

The decisive transition occurred with the addition of the last layer. One reviewing run downloaded the raw \texttt{Fe.win} file from Tutorial~18 and found \texttt{dis\_froz\_max~$=$~30} but rationalized the contradiction away as pseudopotential-specific and confirmed the original finding with caveats. Adding the instruction ``A finding is either completely correct or it is not. There is no middle ground'' removed that escape hatch. Subsequent runs then independently deprecated the target finding.

\paragraph{Delivery gap: recipe replication anti-pattern.}
The delivery experiment asked whether the corrected 21-insight Fe database would alter the behaviour of a fresh Fe calculation agent. It did not. Across six delivery runs, every agent still chose \texttt{dis\_froz\_max~$=$~17}~eV. By runs 19--23 the corrected finding was visible; what failed was the comparative, claim-testing cognitive mode required to interpret it. One run provided the clearest evidence:

\begin{quote}\itshape
``I'm reconsidering the \texttt{dis\_froz\_max} value --- the knowledge base shows that 17~eV gives $-921$~S/cm without adaptive refinement and $-771.5$~S/cm with it, while the official tutorial uses 30~eV\ldots{} Based on the adaptive refinement results, \texttt{dis\_froz\_max~$=$~17}~eV is the better choice since it produced the strongest conductivity value of $-771.5$~S/cm.''
\end{quote}

This is the same error-cancellation trap the review agent learned to catch. The calculation agent compressed the retrieved evidence into a recipe-oriented summary and anchored on the best prior numerical outcome rather than on parameter convergence. The delivery failure is therefore not merely a search-ranking problem; it is evidence that in execution mode, the claim-testing cognitive mode required to interpret contradictory knowledge entries fails even when the corrected entries are visibly retrieved. This recipe-replication behaviour is the same anti-pattern observed in the Ni cross-material experiment in the main text.

\section{Benchmark statistics}
\label{app:bench}

\paragraph{Category-resolved statistics for the 135-material matrix.}
\begin{center}
\small
\setlength{\tabcolsep}{4pt}
\begin{tabular}{@{}lcccccc@{}}
\toprule
Category & Materials & Complete & Success (\%) & Lattice MAE (\%) & Band-gap MAE (\%) & Mean insights \\
\midrule
Elemental & 31 & 31 & 100.0 & 0.77 & 24.3 & 2.6 \\
Semiconductor & 35 & 32 & 91.4 & 1.43 & 51.1 & 2.7 \\
Oxide & 25 & 19 & 76.0 & 0.66 & 41.5 & 3.5 \\
Halide & 13 & 13 & 100.0 & 1.23 & 38.8 & 3.1 \\
TMC & 23 & 15 & 65.2 & 0.44 & 28.4 & 2.8 \\
Topological & 8 & 5 & 62.5 & 2.50 & 61.7 & 2.6 \\
\midrule
All & 135 & 115 & 85.2 & 1.02 & 46.4 & 2.9 \\
\bottomrule
\end{tabular}
\end{center}

All 20 failures were timeouts under a uniform end-to-end session budget; no wrong-answer completions and no unrecovered engine crashes occurred in the final statistics. The ORCA benchmark yielded 91/98 completed under the conservative publication criterion; the reference-backed raw-ORCA-converged subset achieved bond-length MAE $=$ 0.0069~\AA{} (0.52\%) across 78 comparisons and bond-angle MAE $=$ 0.51$^\circ$ across 33 comparisons.

\paragraph{Tool usage across the 135-material benchmark.}
\begin{center}
\small
\begin{tabular}{@{}lcc@{}}
\toprule
Tool & Total calls & Per material \\
\midrule
\texttt{inspect\_calculation} & 603 & 5.2 \\
\texttt{set\_parameters} & 485 & 4.2 \\
\texttt{search\_knowledge} & 480 & 4.2 \\
\texttt{get\_results\_summary} & 400 & 3.5 \\
\texttt{run\_calculation} & 386 & 3.4 \\
\texttt{search\_demos} & 382 & 3.3 \\
\texttt{record\_insight} & 371 & 3.2 \\
\bottomrule
\end{tabular}
\end{center}

Two behavioural features follow directly. Knowledge retrieval was systematic rather than exceptional (\texttt{search\_knowledge} used 4.2 times per material). Knowledge recording was likewise routine at the finding level: the median completed material generated three recorded insights.

\paragraph{Nudge compliance and the bandwidth problem.}
Across 398 execution sessions, 338 (84.9\%) received at least one recording nudge, yet only 16 sessions (4.0\%) contained any mid-session \texttt{record\_insight} call, 95.3\% of all recordings occurred in the final 10\% of session events, and zero execution-mode sessions produced pattern-grade entries. All 41 patterns in the knowledge base were created in dedicated reflection sessions rather than during active calculations. Agents often recognized cross-compound trends informally during calculations, but execution mode did not provide enough attentional slack to convert those observations into higher-grade, provenance-linked knowledge objects.

\section{Dedicated reflection distills findings into predictive knowledge}
\label{app:reflection}

\paragraph{Two independent reflection sessions converged on the same three patterns.}
The dedicated distillation experiment asked whether a short reflection session could extract higher-order structure from 25 findings accumulated across 24 execution sessions over 19 zinc-blende semiconductors. Two independent sessions were run on the same starting database with the same 12-word prompt: one used Claude Sonnet 4.6 (2~min 46~s), one used Claude Opus 4.6 (2~min 28~s). Both produced exactly three pattern-grade entries with identical scientific themes:

\begin{itemize}
\item PBE lattice-constant overestimation increases systematically with constituent atomic mass across the tetrahedral semiconductor families.
\item PBE band-gap underestimation worsens as the true gap narrows and becomes qualitatively unreliable in the near-zero-gap regime.
\item Variable-cell relaxations with semicore pseudopotentials require higher cutoffs and smearing to avoid Pulay-stress artifacts.
\end{itemize}

The wording differed, but the numerical backbone was the same (e.g., the monotonic Group IV sequence C~$+$0.14\%, Si~$+$0.71\%, Ge~$+$1.80\%, $\alpha$-Sn~$+$2.58\%). The reflection sessions were not prompted with target answers or suggested categories. Pattern structure appears to have been latent in the findings themselves; dedicated review simply exposed it.

\paragraph{Pattern-guided design produced an accurate quantitative prediction for InP.}
The follow-up session tested whether distilled patterns could drive new scientific reasoning. Starting from the Opus-generated database of 25 findings plus 3 patterns, the agent was asked to propose and execute one calculation testing a specific prediction. It selected InP (relaxed structure existed; no band structure yet), used the distilled band-gap pattern to predict a PBE direct gap of $\sim$0.7~eV (corresponding to $\sim$45--48\% underestimation relative to the 1.344~eV experimental direct gap), and used the workflow pattern to choose \texttt{ecutwfc~$=$~80}~Ry, SSSP precision pseudopotentials, and a $12\times 12\times 12$ mesh. The calculation returned a direct band gap of 0.696~eV at $\Gamma$ --- magnitude prediction accurate to 0.6\%, with correct direct-gap character. This is the strongest evidence that pattern-grade knowledge functions as a predictive scientific model rather than a descriptive summary.

\paragraph{Implications.}
The execution sessions generated the empirical substrate. Dedicated review sessions compressed that substrate into higher-order, provenance-linked patterns in under three minutes. The design session then used one of those patterns to guide an efficient new calculation. The full progression from observation to pattern to prediction was reproducible from recorded data and robust across two different models.

\paragraph{Availability.}
Complete experimental artifacts, including full prompts, agent traces, knowledge database snapshots, and reviewed-insight tables, are archived on Zenodo at \url{https://doi.org/10.5281/zenodo.20588420} under CC-BY 4.0.

\begin{figure}[h]
\centering
\includegraphics[width=\textwidth]{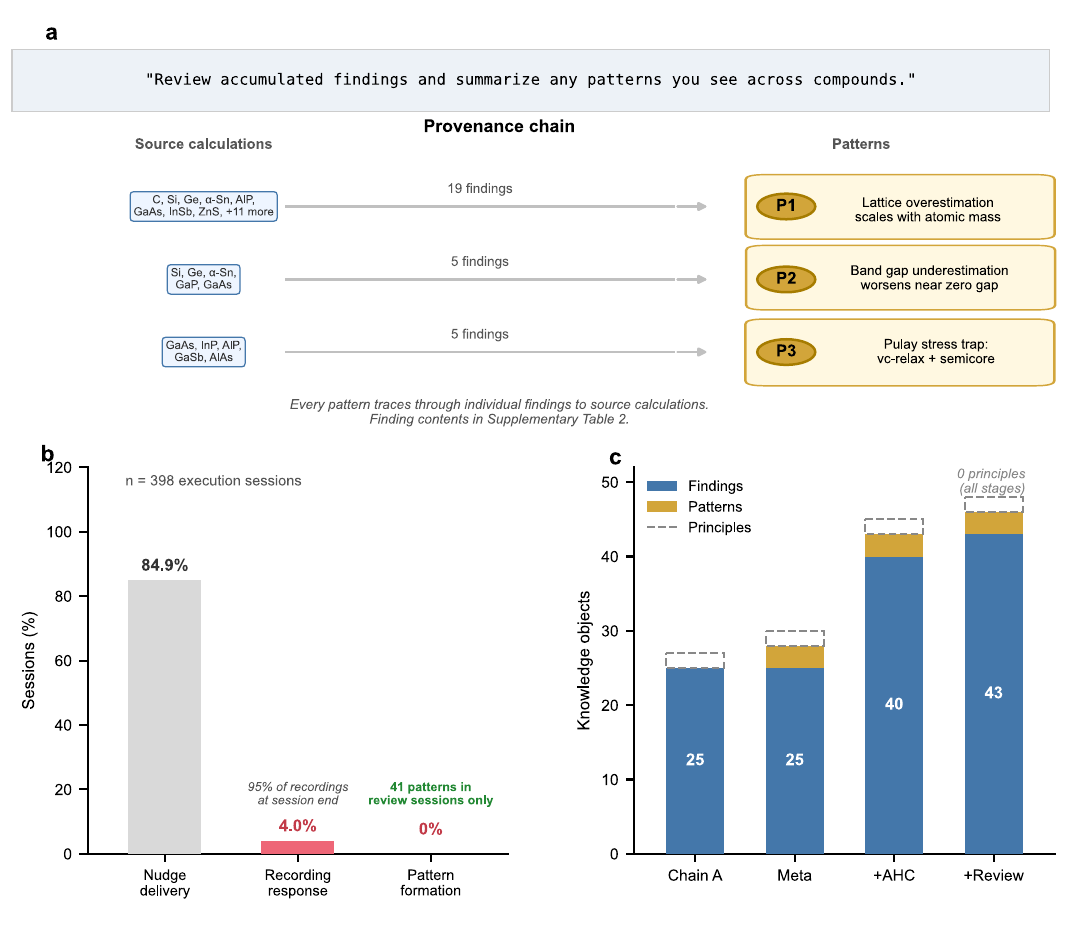}
\caption{\textbf{Knowledge consolidates through dedicated reflection.}
\textbf{a}, Three patterns distilled from 25 findings by a dedicated reflection session (Pattern 1: lattice overestimation scaling; Pattern 2: band gap underestimation; Pattern 3: Pulay stress trap), with provenance chains linking each pattern to its source calculations.
\textbf{b}, Nudge compliance across 398 execution sessions: 84.9\% received recording nudges, 4.0\% showed any mid-session recording, 0\% produced patterns during execution. All 41 patterns were produced exclusively in dedicated review sessions.
\textbf{c}, Knowledge grade evolution across experiments: findings accumulate during execution (25$\to$43), patterns appear only after dedicated reflection (0$\to$3), principles remain an open frontier.
}
\label{fig:consolidation}
\end{figure}

\begin{figure}[h]
\centering
\includegraphics[width=0.9\textwidth]{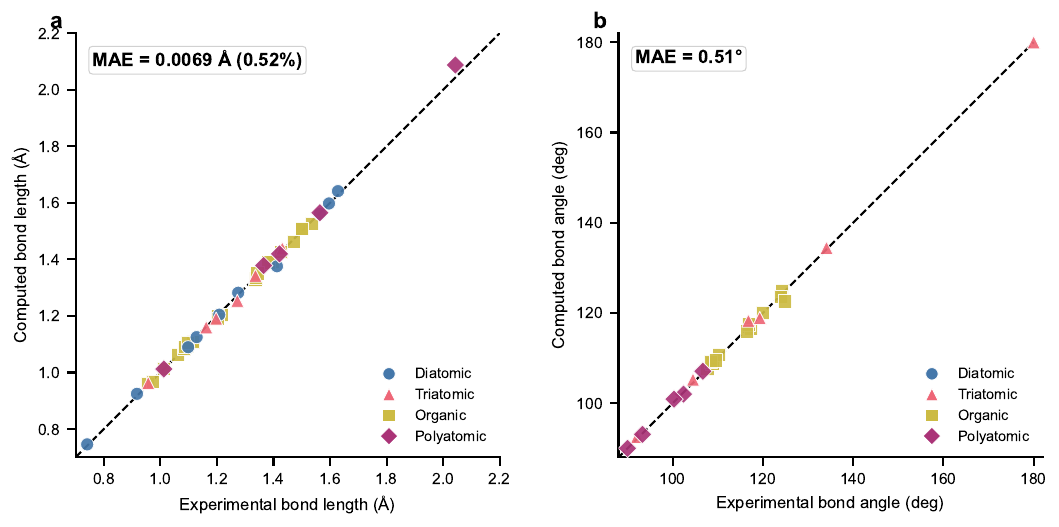}
\caption{\textbf{Cross-engine, cross-model validation: molecular geometry optimization with ORCA 6.1.1 and GPT 5.4.}
\textbf{a}, Computed versus experimental bond lengths for 78 comparisons from the reference-backed raw-ORCA-converged subset (MAE $=$ 0.0069~\AA, 0.52\%), consistent with established hybrid-DFT benchmarks.
\textbf{b}, Computed versus experimental bond angles for 33 comparisons (MAE $=$ 0.51$^\circ$).
This experiment uses a different simulation engine (ORCA vs.\ Quantum ESPRESSO), a different AI model (GPT 5.4 vs.\ Claude), and a different chemical domain (molecular vs.\ solid-state), confirming that the platform generalizes across all three axes. Of 98 molecules, 91 completed under the conservative publication criterion; the seven final failures comprise two agent-declared failures despite converged raw outputs, three missing or non-terminated last-output cases, and two explicit last-attempt error terminations.}
\label{edfig:orca}
\end{figure}

\end{document}